\documentclass[aps,pra,amsfonts,amssymb,epsfig,twocolumn,groupedaddress,showpacs]{revtex4}
\usepackage{graphics}
\usepackage{epsfig}
\usepackage{amsmath}
\begin{document}
\title{Bosonic Memory Channels}
\author{Vittorio Giovannetti$^{1}$}
\author{Stefano Mancini$^{2}$}
\affiliation{$^{1}$NEST-INFM \& Scuola Normale Superiore, I-56126 Pisa, Italy.\\
$^{2}$Dipartimento di Fisica, Universit\`{a} di Camerino, I-62032 Camerino, Italy.}
\date{\today}
\begin{abstract}
We discuss a Bosonic channel model with memory effects.
It relies on a multi-mode squeezed (entangled) environment's state. 
The case of lossy Bosonic channels is analyzed in detail. 
We show that in the absence of input energy constraints the
memory channels are equivalent to their memoryless counterparts. 
In the case of input energy constraint we provide lower and upper 
bounds for the memory channel capacity.
\end{abstract}
\maketitle
Quantum communication with continuous alphabet provides an interesting alternative to 
the traditional discrete alphabet based approach (using e.g. qubits)~\cite{bp01}. 
Much effort has been recently devoted to characterize continuous alphabet quantum 
channels in terms of information capacities~\cite{bs98}. 
At present capacity  results are only known for a restricted class of channels. 
Among them the lossy Bosonic channel, which consists of a collection of Bosonic 
modes that lose energy en route from the transmitter to the receiver.
The classical capacity of 
such communication lines
has been found to be unaffected by the use of entanglend inputs 
among different channel uses
~\cite{getalPRL04},
in close analogy to what happens for a wide class 
of qubit alphabet channels~\cite{kr01}.
However, for the latter it was argued that entangled inputs may enhance the information
transmission in the presence of correlated noise (\textit{memory})~\cite{mp02}. 
Increasing attention has been dedicated to memory effects in quantum channels, 
but only within discrete alphabets \cite{bm04,bdb03}. 

Here, we investigate the memory effects in continuous alphabet channels. 
In Sec.~\ref{s:sezione1} we 
present 
a class of  Bosonic memory channels which
in the case of $n$ 
unconstrained-inputs, 
is unitarily equivalent to 
its i.d.d.
memoryless~\cite{bs98} counterpart with an input
space $n$ times larger than the single use case.
Unfortunately such equivalence is partially lost when imposing energy 
constraint~\cite{caves94,hw01,sha04} on
the input states of the communication line.
Under such conditions in Sec.~\ref{s:sezione2} we  
supply some bounds for the capacity of these channels.
The papers ends with the conclusions in Sec.~\ref{s:sezione3}.

\section{Model}\label{s:sezione1}

A quantum channel that uses continuous alphabet can be modeled by a Bosonic field mode
whose phase space quadratures enable for continuous 
variable encoding/decoding \cite{caves94}.
On $n$ uses of such a channel we have to consider $n$ independent Bosonic modes,
described by annihilation 
operators $a_k$ for $k=1,\cdots,n$.
\begin{figure}[t]
\begin{center}
\epsfxsize=.8\hsize\leavevmode\epsffile{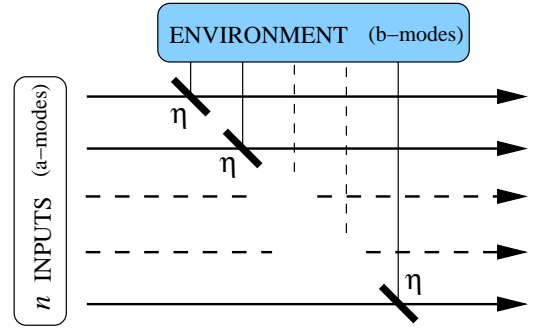}
\end{center}
\caption{(Color online) Scheme of the communication scenario: $n$ uses 
of the lossy Bosonic channel correspond
to $n$ input Bosonic modes $a_k$ interacting with the
environment modes $b_k$ trough $n$ beam splitters.}
\label{f:fig1}
\end{figure}
As depicted in Fig.~\ref{f:fig1} we restrict the analysis
to the case where each $a_k$ interacts with
an environment mode $b_k$ through a 
beam splitter of transmittivity $\eta\in[0,1]$, thus modeling lossy channels. 
The signal-noise coupling is then characterized by 
$U\equiv \otimes_{k=1}^n U_k \label{prima}$
with 
\begin{eqnarray}
U_k=\exp\left[\,(a_k^\dag b_k-a_kb_k^\dag)\arctan\sqrt{\frac{1-\eta}\eta}\;\right]\;
\label{Vdefu}\;,
\end{eqnarray} 
the unitary operator which satisfies the following transformations~\cite{walls94}
\begin{eqnarray}
U_k a_k U_k^\dag &=&\sqrt{\eta}\; a_k - \sqrt{1-\eta} \; b_k\,,
\nonumber\\
U_k b_k U_k^\dag&=&\sqrt{\eta}\; b_k + \sqrt{1-\eta} \; a_k\,.
\label{unouno}
\end{eqnarray}
Let ${r}$ be the density matrix in the Hilbert space 
${\cal H}_{\mbox{\small{tot}}}^{(n)}\equiv \otimes_{k=1}^n {\cal H}_k$ 
which describes the input state of the $n$ channel uses.
Here ${\cal H}_k$ is the Hilbert space associated with input mode $a_k$.
For a memoryless channel the environment acts independently on each
$a_k$. This can be described by assuming the modes $b_k$ to be in the same state $\rho_b$.
The output density matrix corresponding to $r$ is hence given by~\cite{getalPRA04}
\begin{eqnarray}
{\cal L}({r})=\mbox{Tr}_b\left[U\:({r}\otimes {{r}}_b)\:U^\dag\right]\,,
\label{mappanew}
\end{eqnarray}
where the trace is performed over the environment's degrees of freedom, initially in the state
${{r}}_b \equiv \rho_b^{\otimes n}$. Because of the tensorial structure of  
$U$ and ${{r}}_b$, the map~(\ref{mappanew})  becomes
\begin{eqnarray}
{\cal L}({r})= \otimes_{k=1}^n {\ell}_k ({r})\;,
\label{mappanewTEN}
\end{eqnarray}
with $\ell_k$ being the map on ${\cal H}_k$ associated with the $k$-th channel use which transforms
the density matrix  $\rho$ of ${\cal H}_k$  according to
\begin{eqnarray}
\ell_k(\rho)\equiv  \mbox{Tr}_{b_k}\left[U_k\:({\rho}\otimes {\rho}_b)\:U_k^\dag\right]
\;.
\label{Vmappaelle}
\end{eqnarray}

A memory channel is characterized by non trivial correlations between the environment actions
on the different channel uses which cannot be accounted for by Eq.~(\ref{mappanewTEN}).
We model this situation by replacing the separable state $r_b$ of Eq.~(\ref{mappanew})
with the entangled state
\begin{eqnarray}
\tilde{{r}}_b \equiv \Omega_b\; {r}_b \; \Omega_b^{\dag}\,,
\label{mappa1}
\end{eqnarray}
where $\Omega_b$ is a unitary, multi-mode squeezing operator \cite{walls94}
\begin{eqnarray}
\Omega_b \equiv \exp [ \;\sum_{k,k'} ( \xi_{kk'}^*  b_k b_{k'}- \xi_{kk'}
b_k^{\dag} b_{k'}^{\dag})]\,,
\label{squeezing}
\end{eqnarray}
which couples the $b_k$ modes through the squeezing parameters $\xi_{kk'}$. 
The corresponding output state of the channel is hence described by the map
\begin{eqnarray}
\tilde{\cal L}({r})=\mbox{Tr}_b\left[U\:({r}\otimes \tilde{{r}}_b)\:U^\dag\right]\,.
\label{mappa}
\end{eqnarray}
The dependence of Eq.(\ref{mappa}) on $n$ is generally
more involved with respect to that of Eq.~(\ref{mappanewTEN}). 
Equation (\ref{mappa}) also depends on
the parameters $\xi_{kk'}$ and for
for $\xi_{kk'}=0$ it is $\Omega_b=\openone$ and $\tilde{\cal L}={\cal L}$.
It is worth noting that in defining  the memory channel model~(\ref{mappa}) 
it is not necessary to assume Eqs.~(\ref{Vdefu}) and (\ref{squeezing}).
As a matter of fact, $U_k$ can be any unitary operator that couples
the $k$-th channel use mode with its noise $b_k$, 
while $\Omega_b$ can be any unitary operator
which introduces correlations between the $b_k$. 
We will focus on the
case described by Eqs.~(\ref{Vdefu}) and (\ref{squeezing})
since here an interesting simplification occurs. 

\begin{figure}[t]
\begin{center}
\epsfxsize=.99\hsize
\leavevmode\epsffile{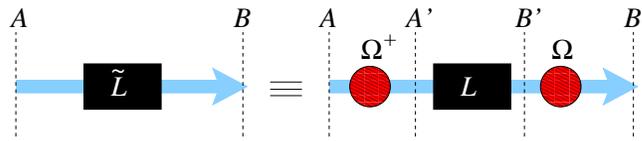}
\end{center}
\caption{(Color online) Decomposition of $\tilde{\cal L}$ of Eq.~(\ref{mappa}).
Input states enter the system in $A$ (input) and leave it in $B$ (output).
According to Eq.~(\ref{fin}) we can identify two intermediate steps: in $A^\prime$ 
the input state has been transformed by 
the unitary operator $\Omega^\dag$ and
enters the map $\cal L$;
in $B^\prime$ it is finally transformed by  $\Omega$.}
\label{f:fig2}
\end{figure}

Our aim is now to relate the memory channel of Eq.~(\ref{mappa}) 
to the memoryless channel
of Eq.~(\ref{mappanew}).
Let us consider
\begin{eqnarray}
\Omega \equiv \exp[ \;\sum_{k,k'} ( \xi_{kk'}^* a_k a_{k'}- \xi_{kk'}
a_k^{\dag} a_{k'}^{\dag})]\,,\label{squeezing1}
\end{eqnarray}
which represents a multi-mode-squeezing 
(unitary) operator acting on the inputs mode $a_k$ with
the same squeezing parameters $\xi_{kk'}$ of~(\ref{squeezing}).
Defining the density matrix
\begin{eqnarray}
\tilde{{r}} \equiv \Omega^{\dag}  {r} \; \Omega\,,
\label{squeezing2}
\end{eqnarray}
and using Eq.~(\ref{mappa1})
we rewrite Eq.~(\ref{mappa}) as
\begin{eqnarray}
\tilde{\cal L}({r})=
\mbox{Tr}_b\left[U  \; (\Omega \otimes \Omega_b )  
\:( \tilde{{r}}\otimes {r}_b ) \: (\Omega^\dag \otimes \Omega_b^{\dag})\;
U^\dag\right]\,.
\label{mappanewintermedio}
\end{eqnarray}
The transformations~(\ref{unouno}) can be used 
to verify that 
\begin{eqnarray}
U \big( a_k a_{k'}  + b_kb_{k'} \big) U^\dag = \; a_k a_{k'} +  b_k b_{k'} 
\;\label{Vomegann} \;,
\end{eqnarray}
which shows that $\Omega \otimes \Omega_b$ commutes with $U$.
Therefore Eq.~(\ref{mappanewintermedio}) yields  
\begin{eqnarray}
\tilde{\cal L}({r})&=&
\Omega \: \mbox{Tr}_b\left[  \;   
U \:(\tilde{{r}}\otimes {r}_b) \: U^\dag  \;\right]\; \Omega^\dag \;,
\label{mappanew1}
\end{eqnarray}
where: {\em i)} since $\Omega$ does not act on $b_k$, 
we have moved it out of the trace operation,
{\em ii)} since $\Omega_b$ 
is unitary we have used the cyclicity of the trace to eliminate it.
Notice that, apart from the unitary operator  $\Omega$, the right-hand side
of Eq.~(\ref{mappanew1}) 
is a standard memoryless Bosonic channel~(\ref{mappanew}) 
which couples the input state $\tilde{{r}}$ with
the environment state ${r}_b$; thus we can write Eq.~(\ref{mappanew1}) as 
\begin{eqnarray}
\tilde {\cal L}({{r}})&=&  \Omega \; {\cal L}(\tilde{{r}}) \;  \Omega^\dag \;.
\label{fin}
\end{eqnarray}
This equation shows that the map 
$\tilde{\cal L}$ can be decomposed in the following
three operations (see also Fig.~\ref{f:fig2}): 
\begin{enumerate}
\item[{\em 1)}]apply the anti-squeezing operator $\Omega^\dag$ to the input state ${r}$;
\item[{\em 2)}]send the resulting state in the channel~$\cal L$; 
\item[{\em 3)}]squeeze the final state with $\Omega$.
\end{enumerate}
Notice that, if the noise parameters $\xi_{kk\prime}$ are known to the
communicating parties, the 
unitary operators $\Omega^\dag$
and $\Omega$ at  {\em 1)} and {\em 3)} 
can always be included in the encoding and decoding stages of
the transmission. In this sense, $\tilde{\cal L}$ and
${\cal L}$ are unitarily equivalent and one expects their ability
in transferring information (classical or quantum) to be the same.

\section{Constrained inputs}\label{s:sezione2}

The ``equivalence'' of $\tilde{\cal L}$ and $\cal L$ 
is partially broken in the case of constrained inputs \cite{caves94}. 
However also in this case,  Eq.~(\ref{fin})  can be used to
relate the capacities of these two channels.
Let us consider for example the capacity of the memoryless channel when 
$\rho_b$'s of Eq.~(\ref{mappanew})
are thermal states with average photon number
$M$, i.e.
\begin{eqnarray}
\rho_b\equiv\frac{1}{M+1}\left(\frac{M}{M+1}
\right)^{b_k^\dag b_k}\,.
\label{vacuum}
\end{eqnarray} 
Under the hypothesis that the inputs states ${r}$ of the channel $\cal L$  
have less then $N$ photons
per channel use, 
\begin{eqnarray}
\mbox{Tr} [\; {r}\; \sum_{k=1}^n a_k^\dag a_k ] \leqslant n N\,,
\label{vacuum1}
\end{eqnarray}
it is believed~\cite{hw01,getalPRL04,sha04} 
that the classical capacity $C({\cal L},N)$ of ${\cal L}$ can be 
saturated by using Gaussian encodings. These allow one 
to achieve a transmission
rate equal to
\begin{eqnarray}
G({\cal L},N) = n \, \left[ \;g(\eta N + (1-\eta) M) - g((1-\eta)M)
\;\right],
\label{capacity}
\end{eqnarray}
where 
\begin{eqnarray}
g(x) = (x+1)\ln(x+1) -x\ln x \;,
\label{Vfunzioneg}
\end{eqnarray} 
and where the linear dependence on $n$ is a consequence
of the absence of memory effects in the transmission.
Even though the identity  $C({\cal L},N)= G({\cal L},N)$
has been proved~\cite{getalPRL04} only for
$M=0$ (environment's vacuum state),
there are strong evidences that it should also
apply for $M>0$. 

\subsection{Upper bounds}\label{s:subsection21}
In the following we derive two independent upper bounds for the
maximum number of classical information  $C(\tilde{\cal L},N)$ 
that can be reliably transmitted 
through the $n$ uses of the memory channel $\tilde{\cal L}$ 
when its inputs $r$ are constrained 
by Eq.~(\ref{vacuum1}).

Equation~(\ref{fin}) establishes that
transmitting ${r}$ into $\tilde{\cal L}$
is equivalent to transmitting $\tilde {r}$ of Eq.~(\ref{squeezing2})
into $\cal L$. The maximum 
average photon number $\overline{N}$ 
per channel use associated with the latter state can be computed using the 
transformations (\ref{unouno}). In particular, for $r$ satisfying Eq.~(\ref{vacuum1}) one
can show that
\begin{eqnarray}
\mbox{Tr} [ \; {\tilde {r}} \; \sum_{k=1}^n a_k^\dag a_k ] \leqslant n \overline{N}\,,
\label{ult}
\end{eqnarray}
where
\begin{eqnarray}
\overline{N} = N \;\left[ \; \cosh(4 \overline{d}) + \sinh(4 |\overline{d}|) \;\right]
+ s_1 + s_2 \geqslant N\,.
\label{NBAR}
\end{eqnarray}
In the above expression 
$s_1$ and $s_2$  are positive quantities defined in Appendix~\ref{s:appendice1}
and $\overline{d}$ is the eigenvalue of the of the $n\times n$ matrix $\xi_{k k'}$
(assumed real symmetric for the sake of simplicity) 
having maximum absolute value.
The quantity $\overline{N}$ 
determines the maximum value of average photon number
per channel use that is entering the channel $\cal L$ at point $A^\prime$ of
Fig.~\ref{f:fig2} when we feed the channel 
$\tilde{\cal L}$ with $N$ photons per use.
We can exploit this fact to conclude that the capacity $C(\tilde{\cal L},N)$ cannot
be greater than the capacity $C({\cal L},\overline{N})$ of the memoryless 
channel ${\cal L}$ with $\overline{N}$ average photon number per channel use, i.e.
\begin{eqnarray}
C(\tilde{\cal L},N) & \leqslant& C({\cal L},\overline{N})
\label{capacitylast} \;.
\end{eqnarray}
Clearly this inequality does not depend on the validity of the
 conjecture~\cite{hw01,getalPRL04,sha04}. However, in order
to derive an explicit expression for the bound~(\ref{capacitylast})  
it is useful to assume~\cite{hw01,getalPRL04,sha04} and evaluating  the right-hand side
term of~(\ref{capacitylast}) by means of the function $G(\tilde{\cal L},N)$
of Eq.~(\ref{capacity}), i.e.
\begin{eqnarray}
C(\tilde{\cal L},N) & \leqslant& n \, 
\left[g(\eta \overline{N} + (1-\eta) M) - g((1-\eta)M)
\right] \;. \nonumber \\
\label{Vcapacitylast}
\end{eqnarray}

An alternative upper bound for $C(\tilde{\cal L},N)$ 
can be obtained by fixing $n$ and by
assuming  the corresponding map $\tilde{\cal L}$ to
represent a memoryless channel.
This allows us to 
derive the following inequality \cite{nota1}
\begin{eqnarray}
C(\tilde{\cal L},N) \leqslant
\sup_m \; C_m(\tilde{\cal L}^{\otimes m},N)/ m
\;, \label{add0}
\end{eqnarray}
where $m$ is the number of successive uses of the ``memoryless''
channel $\tilde{\cal L}$ and where~\cite{HSW}
\begin{eqnarray}
C_m(\tilde{\cal L}^{\otimes m},N) 
&\equiv&  {\max_{p^{(i)},{R}^{(i)} ; N}}\Big\{
 S(\tilde{\cal L}^{\otimes m} ({R})) \label{add} \\
&& \qquad -\sum_j\;p^{(i)} 
S(\tilde{\cal L}^{\otimes m}
({R}^{ (i)})) \; \Big\}  \nonumber \;,
\end{eqnarray} 
is the maximum amount of information the two communicating
parties can share by feeding with 
probabilities $p^{(i)}$ the $m$ copies of $\tilde{\cal L}$ 
with messages ${R}^{(i)}\in \left( {\cal H}^{(n)}_{\mbox{\small tot}} 
\right)^{\otimes m}$.
Here $S({R}) =-\mbox{Tr} [ {R} \ln {R} ]$ is the von
Neumann entropy
and ${R}=\sum_i \;p^{(i)} {R}^{(i)}$ 
is the average input of $\tilde{\cal L}^{\otimes m}$. 
The maximization in Eq.~(\ref{add}) is performed over all ensembles 
$\{ p^{(i)}, {R}^{(i)}\}$ which, for each $\tilde{\cal L}$, satisfy the energy 
constraint~(\ref{vacuum1}), i.e.
\begin{eqnarray}
\mbox{Tr}[ {R} \; ( \sum_{k=1}^n a_{k}^\dag a_{k} )^{\otimes m} ] \leqslant m\;  n N\,.
\label{vacuumEMME}
\end{eqnarray}

Likewise Refs.~\cite{getalPRL04,sha04} we provide an
upper bound for~(\ref{add}) by replacing the first/second term at the
right-hand side with the maximum/minimum respectively 
\begin{eqnarray}
C_m(\tilde{\cal L},N) &\leqslant& m \; \max_{{r} ; \;  N } \Big\{
 S(\tilde{\cal L}({r})) \Big\} 
- \min_{ {R} }
\Big\{ S(\tilde{\cal L}^{\otimes m}({R})) \Big\} .
\nonumber \\ \label{add1}
\end{eqnarray}
The subaddittivity
of the von Neumann entropy has been used to transform the maximization
over ${R}\in \left( {\cal H}_{\mbox{\small tot}}^{(n)}\right)^{\otimes 
m}$ into 
maximization over inputs ${r}$ of ${\cal H}_{\mbox{\small tot}}^{(n)}$, 
and the constraint~(\ref{vacuumEMME}) has been dropped in the 
minimization.
Equation~(\ref{add1}) establishes that $C_m(\tilde{\cal L},N)$
can be bounded by the difference between 
the {\em maximum} output entropy of {\em single} use ($m=1$) of $\tilde{\cal L}$ and
the {\em minimum} output entropy of the $m$ channel uses: let us compute these quantities.

For inputs $r$ 
that satisfy the constraint~(\ref{vacuum1}), Eqs.~(\ref{unouno}) and  (\ref{mappa}) 
establish that the maximum 
average photon number  we can get at the output of the channel $\tilde{\cal L}$ is equal to
$n N_{\mbox{\small out}}$ where,
\begin{eqnarray}
N_{\mbox{\small{out}}}&=& \eta  N +  (1-\eta) ( s_0 M + s_1 )  \;,
\label{energiaOUT}
\end{eqnarray} 
with $s_1$ as in Eq.~(\ref{NBAR}) and
\begin{eqnarray}
s_0 \equiv \sum_{j=1}^{n} \cosh(4 d_j)/ n \geqslant 1\;,\label{esse0}
\end{eqnarray}
(see Appendix ~\ref{s:appendice2} for details).
We can hence upper bound the output entropy of $\tilde{\cal L}$ with 
$n$ times the entropy $g(N_{\mbox{\small{out}}})$ of a 
thermal state whose total average photon number is equal to 
$N_{\mbox{\small{out}}}$ \cite{getalPRL04} .

To compute the minimum output entropy of the channel $\tilde{\cal L}$ we
use a conjecture proposed in Ref. \cite{getalPRA04}.
In fact, from Eq.~(\ref{fin}) and the invariance of $S$ under unitary operations, we have
\begin{eqnarray}
\min_{R} \Big\{ S(\tilde{\cal L}^{\otimes m} (R)) 
\; \Big\}  
= \min_{R} \Big\{ S( {\cal L}^{\otimes m} (R) ) \; \Big\}  \;
\label{mini1}
\end{eqnarray}
According to the analysis of Ref.~\cite{getalPRA04} 
the minimum output entropy of the
channel $\cal L$ should be provided by vacuum input: this result has not
been proven yet but, as in the case of the conjecture Eq.~(\ref{capacity}),
there is strong evidence in support of it (as a matter of fact these 
two conjectures are strongly related). Assuming the
conjecture of Ref.~\cite{getalPRA04}  we can simplify Eq.~(\ref{mini1}) as
follows,
\begin{eqnarray}
\min_{R} \Big\{ S(\tilde{\cal L}^{\otimes m} (R)) \; \Big\} 
= m \; n \; g((1-\eta) M)
\label{mini}\;.
\end{eqnarray}
which replaced in Eq.~(\ref{add1}) and (\ref{add0}) gives,
\begin{eqnarray}
C(\tilde{\cal L},N) &\leqslant&   n \; \big[ \; g \left( \eta \; N +  (1-\eta) \; 
( s_0 M + s_1 ) \right)  \nonumber \\
&& \qquad  - g((1-\eta) M) \; \big]
\label{add2}\; .
\end{eqnarray}
The right-hand sides of Eqs.~(\ref{Vcapacitylast}) and (\ref{add2})
are two independent upper bounds for the capacity of the $n$ successive 
uses of the memory channel $\tilde{\cal L}$. 
They have been derived by assuming the conjectures discussed in 
Refs.~\cite{hw01,getalPRL04,sha04} and Ref.~\cite{getalPRA04},
respectively.
Both of them are greater or equal to
the alleged capacity $G({\cal L},{N})$ 
of Eq.~(\ref{capacity})  of a memoryless channel $\cal L$ with average 
constraint $N$ (it follows for instance from the fact that $g(x)$ is an increasing function of
$x$). 

\subsection{Lower bound}\label{s:subsection22}

A lower bound for $C(\tilde{\cal L},N)$ can be obtained by providing an 
encoding-decoding procedure that allows to achieve reliable information transfer.
This is not a simple task for a memory channel. 
However we can use the decomposition rule~(\ref{fin}) to transform encodings 
of $\cal L$ (which are simpler to characterize) into encodings of $\tilde{\cal L}$.

The only known encoding that allows the memoryless channel $\cal L$ to
asymptotically achieve the transmission rate~(\ref{capacity}) 
requires the sender to feed the channel
with thermal states \cite{getalPRL04,hw01,sha04}.
Suppose that she/he manages to produce a thermal state 
at point $A^\prime$ of Fig.~\ref{f:fig2} and assume that the average photon
number of such state is $N^\prime$.
This means that the state of the $n$ modes in $A^\prime$ is given by
\begin{eqnarray}
\tilde r = \bigotimes_{k=1}^n \frac{1}{N^\prime+1} \left( \frac{N^\prime}{N^\prime+1} 
\right)^{a^\dag_k a_k}\,.
\label{thermalAprime}
\end{eqnarray}
The corresponding state in $A$ is obtained by inverting the
relation Eq.~(\ref{squeezing2}) and has average photon number equal to 
\begin{eqnarray}
\mbox{Tr} [ \; \Omega \; \tilde{r} \; \Omega^\dag  \sum_k a_k^\dag a_k ] = n \; 
( s_0\; N^\prime + s_1)\,,
\label{finale} 
\end{eqnarray}
with $s_0$ and $s_1$ as in Eqs.~(\ref{NBAR}) and (\ref{energiaOUT})
(see Appendix~\ref{s:appendice3} for details).
Since we are allowed to supply less then $N$ average
photon number per channel use, we should require
\begin{eqnarray}
s_0\; N^\prime + s_1 \leqslant N\,, \qquad \Longrightarrow 
\qquad N^\prime \leqslant \frac{N - s_1}{s_0} \leqslant N\,.
\label{finale1}
\end{eqnarray}
For all $N^\prime$ satisfying the above relation the sender is able to
use the optimal encoding~(\ref{thermalAprime}) to transfer messages with
capacity $G({\cal L},N^\prime)$ given in Eq.~(\ref{capacity}). 
This means that for 
large enough $n$, 
the following inequality holds
\begin{eqnarray}
&&C(\tilde{\cal L},N)  \geqslant  G({\cal L},  (N - s_1)/{s_0}) \;.
\label{capacitylast11} 
\end{eqnarray} 
Since $C({\cal L},  N)$ is always greater than the right-hand side
of Eq.~(\ref{capacitylast11}), we cannot claim that
$C(\tilde{\cal L},N)$ is definitely greater than  $C({\cal L},N)$.

\section{Conclusions}\label{s:sezione3}

We have discussed a model of quantum \textit{memory} 
channel employing continuous alphabets which relies on the use of multi-mode squeezed 
(entangled) environment state. 
In the simple case of lossy Bosonic channel we have found a unitarily equivalence~(\ref{fin}) 
between the map $\tilde{\cal L}$ of the memory channel and the map $\cal L$ of its 
memoryless counterpart. When no constraints on the input states there is
a perfect equivalence in the ability of the these channels in transferring
information.
As a consequence, entangled inputs can only be used to reach 
an optimal encoding, but they do not improve the channel's performance. 
This shows that the role of 
entanglement is subtle.
In particular, it seems no longer useful when other unlimited resources are 
available.
In the more realistic scenario of energy constrained
input states, we provided upper and lower bounds for the
capacity of the memory channel.
In particular, from Eqs.~(\ref{capacitylast}) and 
(\ref{capacitylast11}) we have
\begin{eqnarray}
G({\cal L},  (N-s_1)/s_0) \leqslant C(\tilde{\cal L},N) \leqslant C({\cal L},\overline{N})\,,
\label{ultima}
\end{eqnarray}
which, assuming the conjecture~\cite{hw01,getalPRL04,sha04},
shows that the classical capacity of the memory channel is bounded by 
classical capacities of the
memoryless channel ${\cal L}$ having different power constraints.
It is worth noticing that, because of Eq.~(\ref{fin}),
the above relation  generalizes also to all the other 
capacities (e.g. quantum capacity, entanglement assisted 
capacity  \cite{bs98}) of $\tilde{\cal L}$
and ${\cal L}$.

Finally, we believe that the results presented here,
though not giving a conclusive 
answer on the usefulness of entanglement versus memory effects,
are deep enough.
Furthermore,
the presented model is fairly general and could be used to study a 
variety of specific and practical situations.
For instance it could be interesting to analyze the case where Eq.(7) 
only connects nearest neighbors modes, that is, each use is only affected by the previous one.
As such this work paves the way for further studies in Bosonic memory channels.

\appendix

\section{}\label{s:appendice}

Define $Z$ the $n\times n$ matrix whose elements are $\xi_{k k^\prime}$ 
of Eq.~(\ref{squeezing1}). Without loss of generality we can choose $Z$ to be symmetric.
For the sake of simplicity, in the following we will also assume $Z$ to be Hermitian.  
If  $Z$ is symmetric and Hermitian it is also real and we can diagonalize it 
by means of a unitary matrix $V$ of real element $v_{kk^\prime}$, i.e.
\begin{eqnarray}
Z = V\cdot D \cdot V^{T} \; \; \Longrightarrow \; \; 
 \xi_{kk^\prime} = \sum_{j=1}^{n} d_j v_{k j} v_{k^\prime j} \;,
\label{diago}
\end{eqnarray}
where $D$ is a $n\times n$ diagonal matrix with
real elements $d_j$.
Using the above relation one can verify that 
\begin{eqnarray}
\sum_{kk^\prime} \xi_{kk^\prime} a_k a_{k^\prime} 
&=& \sum_j d_j c_j c_j \;,
\label{diago1}
\end{eqnarray}
where for all $j =1, \cdots, n$
\begin{eqnarray}
c_j &=& \sum_{k} v_{k j} a_k \;, \label{diago2bis}\\
a_k &=& \sum_{j} v_{kj } c_j \;.
\label{diago2}
\end{eqnarray}
Notice that the operator $c_j$ form 
a set of independent annihilation operators which satisfy the usual commutation relations,
\begin{eqnarray}
[c_j, c_{j^\prime}] = 0  \qquad \qquad 
[c_j, c_{j^\prime}^\dag] = \delta_{j j^\prime}
\label{commu}\;,
\end{eqnarray}
and the identity
\begin{eqnarray}
\sum_j c_j^\dag c_j &=& 
\sum_k a_k^\dag a_{k} \label{somma}\;.
\end{eqnarray}
Replacing the relation (\ref{diago2}) in (\ref{squeezing1}) we get,
\begin{eqnarray}
\Omega &\equiv& \exp\left[ \;\sum_{j} d_j ( \; c_j^2 - (c_j^\dag)^2 \;)\right] 
\nonumber \\
&=&
\bigotimes_{j=1}^n \exp\left[ d_j \; c_j^2 - d_j (c_j^\dag)^2 \;\right]\;.
\label{squeezing1equiv}
\end{eqnarray}
Thus the operator $\Omega$ squeezes the mode $c_j$  as follows,
\begin{eqnarray}
\Omega \; c_j \; \Omega^\dag &=& 
\cosh(2 d_j) \; c_j + \sinh(2 d_j) \; c_j^\dag
\label{squeezingci}\;, \\
\Omega^\dag \; c_j \; \Omega &=& 
\cosh(2 d_j) \; c_j - \sinh(2 d_j) \; c_j^\dag \;.
\nonumber
\end{eqnarray}

\subsection{Derivation of Eq.~(\ref{ult})} \label{s:appendice1}

From Eq.~(\ref{fin}) we know that 
transmitting the state $r$ into $\tilde{\cal L}$
is equivalent to transmitting the state $\tilde r$ of Eq.(\ref{squeezing2})
 into $\cal L$. Equations~(\ref{somma}) and~(\ref{squeezingci}) allow us to compute the average
photon number of $\tilde r$ as follows 
\begin{eqnarray}
&&\mbox{Tr} [ \; \tilde{r} \; \sum_{k=1}^n a_k^\dag a_k ] = 
\mbox{Tr} \left[\;  r \; \Omega \left( \sum_{k=1}^n a_k^\dag a_k \right)  \Omega^\dag \;
\right]  \nonumber 
\\
&&\qquad \qquad =
\mbox{Tr} \left[ \; r \; \Omega \left( \sum_{j=1}^n c_j^\dag c_j \right)  \Omega^\dag \;\right]
\nonumber  \\
&&\quad =
\;\mbox{Tr} \Big[ r \; \Big(  \sum_{j=1}^n 
 \cosh(4d_j)  \; c_{j}^\dag c_j \nonumber \\
&&\quad   +   
\sinh(4d_j)
(c_{j}^\dag c_j^\dag +c_j c_j)/2 \Big) \; \Big] + n \; s_1\label{vacumm11} \;,
\end{eqnarray} 
where
\begin{eqnarray}
s_1 \equiv \sum_{j=1}^{n} \sinh^2(2 d_j)/ n \geqslant 0
\label{SIGMA}\;.
\end{eqnarray}
Equation~(\ref{vacumm11}) can be upper bounded using the following
inequalities,
\begin{eqnarray}
&&\sum_{j=1}^n \cosh(4 d_j)
\;\mbox{Tr} \left[ r \; c_{j}^\dag c_j \right] \leqslant \cosh(4 \overline{d}) 
\;\mbox{Tr} \left[ r \; \sum_{j=1}^n c_{j}^\dag c_j \right]\nonumber 
\\&&=
 \cosh(4 \overline{d}) 
\;\mbox{Tr} \left[ r \; \sum_{k=1}^n  a_{k}^\dag a_k \right] \leqslant n N \cosh(4 \overline{d})
\label{ine}\;,
\end{eqnarray}
where
$\overline d$ is the eigenvalues $d_j$ with maximum absolute value. 
In deriving the above expression we used~(\ref{somma}) and the fact that
$r$ has at most $N$ average photon number per channel uses.
Analogously we have
\begin{eqnarray}
&&\frac{1}{2}\sum_{j=1}^n \sinh(4d_j)
\;\mbox{Tr} \left[ r \; (c_j^\dag c_j^\dag + c_{j} c_j) \right] \nonumber \\ 
&&\leqslant \; \frac{1}{2}  \sum_{j=1}^n \sinh(4 |d_j|)
\;\mbox{Tr} \left[ r \; (2 c_j^\dag c_j + 1 ) \right] \; 
\nonumber \\
&&\leqslant n N  \sinh(4 |\overline{d}|) + n \; s_2
\label{ine2}\;,
\end{eqnarray}
with
\begin{eqnarray}
s_2 = \sum_{j =1}^{n} \sinh(4 |d_j|)/(2n) \geqslant 0
\label{SIGMA1}\;.
\end{eqnarray}
Equation~(\ref{ult}) finally follows by replacing~(\ref{ine}) and (\ref{ine2}) in
(\ref{vacumm11}).

\subsection{Derivation of Eq.~(\ref{finale})} \label{s:appendice3}
Here we compute the average photon number associated with the state 
$\Omega \; \tilde{r} \; \Omega^\dag$ with $\tilde{r}$ the thermal state defined 
in~(\ref{thermalAprime}). We proceed as in Eq.~(\ref{vacumm11}) obtaining
\begin{eqnarray}
&&\mbox{Tr} [ \; \Omega \; \tilde{r} \; \Omega^\dag  \sum_k a_k^\dag a_k ] 
= \;\mbox{Tr} \Big[ \tilde{r} \; \Big(  \sum_{j=1}^n 
 \cosh(4d_j)  \; c_{j}^\dag c_j \nonumber \\
&&\qquad   -   
\sinh(4d_j)
(c_{j}^\dag c_j^\dag +c_j c_j)/2 \Big) \; \Big] + n \; s_1 \nonumber \\
&& \qquad =\sum_{j=1}^n \cosh(4d_j) N^\prime  + n\; s_1
\label{VVVvacumm11} \;.
\end{eqnarray} 
where the last identity follows from~(\ref{diago2bis}) 
and from the properties 
\begin{eqnarray}
\mbox{Tr} [ \; \tilde{r} \;  a_k^\dag a_{k^\prime} ] &=& \delta_{kk^\prime} N^\prime \nonumber
\\ \mbox{Tr} [ \; \tilde{r} \;  a_ka_{k^\prime} ] &=& 0\;.
\label{Vmaster}
\end{eqnarray}

\subsection{Derivation of Eq.~(\ref{energiaOUT})} \label{s:appendice2}

The average photon number at the output of the channel $\tilde{\cal L}$
associated with the input $r$ can be computed using the relation
\begin{eqnarray}
U^\dag \; a^\dag_k a_k \; U 
&=& \eta a^\dag_k a_k + (1-\eta) b^\dag_k b_k \nonumber \\
&& +  \sqrt{\eta (1-\eta)}
( b^\dag_k a_k + a_k^\dag b_k )\;,
\label{Vrelazione}\end{eqnarray}
and the result of the previous section. In particular from the definition~(\ref{mappa})
we have
\begin{eqnarray}
&&\mbox{Tr}_a \left[ \; \tilde{\cal L}(r) \sum_{k=1}^n a^\dag_k a_k \;\right ] =
 \eta \; \mbox{Tr}_{a,b} \left[\; r \otimes \tilde{r}_b\;  
\sum_{k=1}^n a^\dag_k a_k \;  \right] \nonumber \\
&&\qquad + (1-\eta) \; \mbox{Tr}_{a,b} 
\left[\; r \otimes \tilde{r}_b\;  
\sum_{k=1}^n b^\dag_k b_k \;  \right] \label{energia} \\ 
&&\qquad +  \sqrt{\eta (1-\eta)} \; 
\mbox{Tr}_{a,b} \left[\; r \otimes \tilde{r}_b\; \nonumber 
 \sum_{k=1}^n ( b^\dag_k a_k + a_k^\dag b_k ) \;  \right] \;.
\end{eqnarray} 
The first term on the right-hand side is proportional to the input average photon
number of $r$. The second and the third instead can be computed as 
in Eq.~(\ref{VVVvacumm11}).

\acknowledgments
VG gratefully acknowledges useful discussions with N. J. Cerf and for
informing him of his recent work on memory effects in Bosonic channels~\cite{CERF}.
The contribution of VG to this  
work was supported by the European Community under contracts IST-SQUIBIT,
IST-SQUBIT2, and RTN-Nanoscale Dynamics.


\begin{references}
\bibitem{nota1} The right-hand side term of Eq.~(\ref{add0}) is the
	asymptotic limit for $m\rightarrow \infty$  of the ratio between 
	the maximum amount of classical information $C_m(\tilde{\cal L}^{\otimes m},N)$
	that can be reliably transmitted using
	$m$ copies of a channel whose map is $\tilde{\cal L}$. Clearly this
	cannot be smaller than the amount of classical information that one can
	transmit reliably by just using $\tilde{\cal L}$ once.
\end{references}
\end{document}